\documentclass[prl,twocolumn,superscriptaddress,showpacs]{revtex4}
\usepackage{hyperref}
\usepackage{graphicx}
\usepackage{amsmath,bm}
\def\beqra{\begin{eqnarray}} \def\eeqra{\end{eqnarray}}
\def\beqast{\begin{eqnarray*}} \def\eeqast{\end{eqnarray*}}
\def\beq{\begin{equation}}      \def\eeq{\end{equation}}

\def\Ref{\@startsection{section}{1}{\z@}{-3ex plus-1ex minus-.2ex}%
        {2ex plus.2ex}{\large\sc}}

\begin{document} %%%%%%%%%%%%%%%%%%%%%%%%%%%%%%%%%%%%%%%%%%%%%%%%%%

\preprint{UTEXAS-HEP-00-11}
\preprint{SMU-HEP-00-18}

\title[Drell-Hearn]{The Drell-Hearn Sum Rule at Order $\bm{\alpha^3}$}

\author{Duane A. Dicus}
\email{phbd057@utxvms.cc.utexas.edu}
\affiliation{Center for Particle Physics and Department of Physics\\  The
University of Texas at Austin \\  Austin, Texas 78712}
\author{Roberto Vega}
\email{vega@mail.physics.smu.edu}
\affiliation{Department of Physics\\ Southern Methodist University\\
Dallas, Texas 75275 }

\date{\today}

\begin{abstract}
The Drell-Hearn-Gerasimov-Iddings (DHGI) sum rule for
electrons is evaluated at order $\alpha^3$ and shown to agree 
with the Schwinger contribution to the anomalous magnetic moment.
\end{abstract}

\pacs{11.55.Hx, 13.40.Em, 13.60.Fz}

\maketitle

%%\\[12pt]
%\pagebreak
%\baselineskip=18pt
%%\parskip=12pt

The Drell-Hearn-Gerasimov-Iddings (DHGI) sum rule
\cite{1966} 
relates the anomalous magnetic moment of a
particle to the integral of a difference of cross sections
\begin{equation}
\frac{2\pi^2\alpha \kappa^2}{m^2} = \int^{\infty}_0
\,\frac{d\omega}{\omega}\;[\sigma_P(\omega)-\sigma_A(\omega)]
\label{eqn1}
\end{equation}
$\kappa$ is the anomalous moment in units of $e/2m$, $m$ is the particle's mass and
$\sigma_{{\bm P},A} (\omega)$ are the total cross sections for the
scattering of a circularly polarized photon of energy $\omega $ in the
laboratory system with polarization parallel or antiparallel to the
particle's spin.  In the derivation of~(\ref{eqn1}) the polarization flip of the
photon picks out the magnetic interaction and the coefficient goes as 
$\vec\epsilon\times\vec\epsilon^*$ which requires circular
polarization.  A basic, and as yet unproven, assumption in the
derivation of the sum rule is that the difference between the
amplitude for spin parallel and antiparallel obeys an unsubtracted 
dispersion relation.   

The same assumption is also needed to derive
the Adler sum rule\cite{adler}.   Furthermore, the difference of cross
sections in the right hand side
of~(\ref{eqn1}) is, up to an overall factor, the $Q^2\rightarrow 0$
limit of the
integrand in the Bjorken sum rule~\cite{drell,bj}.   Thus applied to QCD the
DHGI sum rule provides a connection between photoproduction and deep
inelastic scattering.   It provides an important constraint which
complements the sum rules in high $Q^2$ polarized deep inelastic
scattering\cite{bass}.   It is worth emphasizing that the
underlying physics for these sum rules is derived from the very general
predictions of a gauge invariant, local, relativistic quantum field
theory.  Therefore, experimental verification of the
Drell-Hearn-Gerasimov-Iddings (DHGI) sum rule as well as the other sum
rules is of fundamental
importance.   The Bjorken sum rule for the iso-vector part $g_1$ 
has been verified in polarized deep inelastic experiments at CERN and SLAC
to within 8\%\cite{SMC}.   In contrast the DHGI is on less firm ground
as far as experiment is concerned\cite{karliner}; contrary to
expectations the photoproduction of pions fails to saturate the sum
rule.  The reasons for this discrepancy are not yet understood.
Fortunately, there are a series of forthcoming or planned experiments
which will directly test the DHGI sum rule\cite{experiments} in QCD. 

The DHGI sum rule also provides a useful tool in testing the standard
model and in the search for physics beyond the standard model.   
To lowest order in QED the sum
rule tells us that particles of any spin, whether composite or not,
must posses a gyromagnetic ratio, $g=2$.   Schwinger's result \cite{1948}
for the anomalous moment in QED, $\kappa = \alpha/2\pi$, implies
that to order $\alpha^2$ all contributions to the integral of 
$\Delta\sigma$ must
vanish. Some time ago Altarelli, Cabbibo, and Maiani\cite{1972} 
verified that, for instance,
\[\int^{\infty}_0\,\dfrac{d\omega}{\omega}\Delta\sigma^{tree}_{\gamma
  e\rightarrow\nu W} =0,\]
only for $g_W=2$.
Using these results Brodsky et.al.\cite{stan1} were able
to determine the position of the radiation zero in
$\gamma e\rightarrow W\nu$ and to investigate the sensitivity of the zero
position to an anomalous trilinear gauge coupling.  

In other applications Brodsky and Drell\cite{sid_stan} showed that if
a lepton (L) had substructure which could be photo-excited above a
threshold $m^*$, then there would be a contribution to the sum rule of order
${m_L}/{m^*}$.  Thus, deviations from the sum rule predictions could
be interpreted in terms of a mass scale $(m^*)$ for substructure.
In reference~\cite{dbv} the DHGI sum rule is used, within the framework of
the Chiral Quark Model in the large $N_{c}$ limit, to compute the
proton and neutron anomalous magnetic moments.   Reasonable agreement
with experiment was obtained.   Finally, the DHGI sum rule has also
been used to test the consistency of certain extra-dimensional models
of quantum gravity\cite{goldberg}.

In view of this recent activity regarding the DHGI sum rule we
undertook the task of checking the sum rule, to order $\alpha^3$, 
in the simplest of models, QED of the electron.  To our knowledge,
this is the first instance in which the lowest order {\it non-zero}
contribution to the right hand of~(\ref{eqn1}) has been
theoretically computed.
We find that the DHGI sum rule is indeed satisfied to order $\alpha^3$.

We now proceed with an outline of the calculation of the right hand side of
(\ref{eqn1}) to order $\alpha^3$.  At this order there are three
possible contributions, pair production,
$\gamma + e \rightarrow e + e + \bar{e}$, the virtual corrections to Compton 
scattering,  and double Compton scattering, 
$\gamma + e \rightarrow \gamma + \gamma + e$.  The
first of these is zero.  We consider them in turn.

\begin{figure}
\centering\includegraphics[scale=.5]{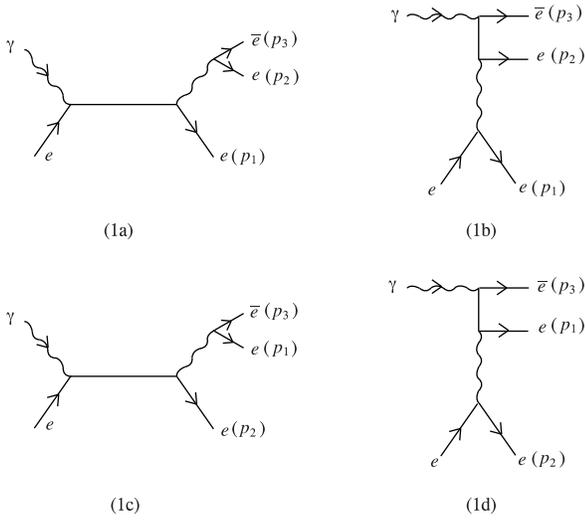}
%\epsfbox{DIfig1}
\caption{\footnotesize 
Examples of the diagrams for $\gamma+e\rightarrow e+e+\overline e$.  The
particles in the final state are labeled with their momenta.  
The diagrams with the
photons crossed are not shown.}
\end{figure}

\section*{\bm{$\gamma + \lowercase{e}\rightarrow\lowercase{e}+
\lowercase{e}+\overline{\lowercase{e}}$}}
A representative set of diagrams for this process is shown in Fig. 1.  The
square of the full set of diagrams of which (1a) and (1b) are examples must
contribute zero to the sum rule.  The upper $e\overline e$ could be
replaced by $\mu\overline\mu$, $\tau\overline\tau$, $ W^+W^-$, etc.
To the order in which we are working the left hand side
of~(\ref{eqn1}) does not involve the masses of any of these particles,
therefore, these contributions must vanish.
We have shown by direct calculation that this is true.  The same is
true for the square of the diagrams represented by ~(1c) plus ~(1d).  
What is not
obvious is that the complete cross terms between diagrams of type ~(1a), (1b)
and those of (1c), (1d) ~must be zero  
%\pagebreak
%\noindent
because this contribution exists only for
electrons. Nevertheless, we have computed these 
cross terms and found that they also vanish.

\section*{\bm{$\gamma+\lowercase{e}\rightarrow\gamma
          +\lowercase{e}~ \lowercase{\rm at~ order}~ \alpha^3$}}
The virtual radiative corrections to unpolarized Compton scattering 
were calculated by Brown
and Feynman\cite{1952} in a well known and classic paper.  Because the sum rule
involves particular spins for the initial particles we need the corrections for
polarized scattering and indeed these have been carefully calculated by Tsai,
DeRaad and Milton\cite{tsai} for every helicity combination.  A
byproduct of our results is that they provide a nice check on the
results reported in reference~\cite{tsai}.
If $\sigma_1,\;\sigma_2$ and $\lambda_1,\;\lambda_2$ label the spins
of the initial
and final electrons and photons and the amplitudes of order $e^n$ are given by
$f^{(n)}(\sigma_2\lambda_2;\sigma_1\lambda_1$) then we need the combination
\begin{multline}
2\,\biggl[f^{(2)} (++;++)f^{(4)}(++;++)\\+f^{(2)}(++;--)f^{(4)}(++;--)  \\
-f^{(2)}(-+;+-)f^{(4)}(-+;+-)\\-f^{(2)}(-+;-+)f^{(4)}(-+;-+)\biggr]. \notag
\end{multline}
These amplitudes are given in Ref. \cite{tsai}
in terms of invariants which can easily be
expressed by $\omega$ and the scattering angle $z$.  Thus the contribution to
(1) involves a double integral over $\omega$ and $z$.
We express the sum rule (1),
with $\kappa$ set equal to $\alpha/2\pi$, as a sum of the virtual and the double
Compton contributions scaled to unity
\begin{equation}  1=I^{(V)}+I^{(DC)}. \label{eqn2}  \end{equation}

It was shown by Feynman and Brown\cite{1952} that the infrared
divergence, which arises as the photon energy approaches zero, cancels
between the two terms in~(\ref{eqn2}).  Since we performed the DHGI
integral numerically, special care had to be taken to ensure an
accurate cancellation.  If we express the virtual contribution as 
\begin{equation}
I^{(V)}=A+B \ln (\omega_{Min}), 
\end{equation}
where $\omega_{Min}$ is the
minimum detectable photon energy (in units of the electron mass), we
find 
\begin{equation}
A=9.68~~,~~B=-4.74.  
\end{equation}
There is an uncertainty in these
numbers due to the numerical integration.  By evaluating (3) for
various $\omega_{Min}$ we estimate these errors to be about 0.03 in
$A$ and 0.01 in $B$.

\section*{ \bm{$\gamma+\lowercase{e}\rightarrow\gamma+\gamma+\lowercase{e}$}}
The contribution to (1) from double Compton scattering is difficult to do
accurately because the difference of cross sections 
decreases slowly even at very
large $\omega$.  Also, in (2), we must cancel numbers on the order of 50 to 100
for $\omega_{Min} = 10^{-4}$ to $10^{-8}$. If we write $I^{DC}$ in (2) as
\begin{equation}
I^{(DC)} = C+D \ln \;(\omega_{Min}),   
\end{equation}
we find,
\begin{equation}
C=-8.70~~,~~D=4.74,  
\end{equation}
with errors in $C$ and $D$ comparable to those of $A$ and $B$.

Thus, within the errors, we conclude that 
\begin{align}
\addtocounter{equation}{1}
B+D&=0 \tag{\theequation a} \\
A+ C&=1 \tag{\theequation b},
\end{align}
and the sum rule (1), as expressed in (2) with $\kappa=\alpha/2\pi$, is verified.

As mentioned above there will be attempts in the near future to
  resolve the discrepancy in the sum rule for nucleons.  It is often 
  mentioned that the convergence of the sum rule would be destroyed by
  the presence of a J=1 fixed pole.  In the language of current algebra
  this translates into an extra term in the commutator of the charge
  densities; it was suggested in Ref~\cite{Chang} that such a term
  could give an additional contribution to the left hand side of (1) 
  that could ameliorate the nucleon discrepancy.  We have shown here 
  that there is no such extra contribution for electrons.   This
  result agrees with those presented in reference~\cite{pantforder}.

 In summary we have completed the first theoretical calculation of
  the DHGI sum rule to order ${\cal O}(\alpha^3)$ in pure QED.  The results 
  presented here support the no-subtraction assumption and affirm the
  validity of the DHGI sum rule.

We would like to thank Scott Willenbrock for reading the manuscript
and providing useful comments, and Hong-Jian He for useful discussions.
This research is funded in part by the Department of Energy under
contracts DE-FG03-95ER40908 and DE-FG03-93ER40757.

% \centerline{\bf Figure 1}
% \vspace{24pt}

% Examples of the diagrams for $\gamma+e\rightarrow e+e+\overline e$.  The
% particles in the final state are labeled with their momenta.  
%The diagrams with the
% photons crossed are not shown.

\end{document}